\documentclass{article}%
\usepackage{amsfonts}
\usepackage{amsmath}
\usepackage{amssymb}
\usepackage{graphicx}%
\begin{document}

\title{Uniformly Accelerated Charge in a Quantum Field: From Radiation Reaction to
Unruh Effect}
\author{Philip R. Johnson\thanks{Electronic address: philipj@physics.umd.edu} \ and B.
L. Hu\thanks{Electronic address: hub@physics.umd.edu}\\\textit{Department of Physics, University of Maryland, College Park, Maryland
20742-4111, USA}}
\maketitle

\begin{abstract}
We present a stochastic theory for the nonequilibrium dynamics of charges
moving in a quantum scalar field based on the worldline influence functional
and the close-time-path (CTP or in-in) coarse-grained effective action method.
We summarize (1) the steps leading to a derivation of a modified
Abraham-Lorentz-Dirac equation whose solutions describe a causal semiclassical
theory free of runaway solutions and without pre-acceleration patholigies, and
(2) the transformation to a stochastic effective action which generates
Abraham-Lorentz-Dirac-Langevin equations depicting the fluctuations of a
particle's worldline around its semiclassical trajectory. We point out the
misconceptions in trying to directly relate radiation reaction to vacuum
fluctuations, and discuss how, in the framework that we have developed, an
array of phenomena, from classical radiation and radiation reaction to the
Unruh effect, are interrelated to each other as manifestations at the
classical, stochastic and quantum levels. Using this method we give a
derivation of the Unruh effect for the spacetime worldline coordinates of an
accelerating charge. Our stochastic particle-field model, which was inspired
by earlier work in cosmological backreaction, can be used as an analog to the
black hole backreaction problem describing the stochastic dynamics of a black
hole event horizon.

\vskip.5cm \noindent{\small \textit{- Invited talk given by BLH at the
International Assembly on Relativistic Dynamics (IARD), June 2004, Saas Fee,
Switzerland. To appear in a special issue of Foundations of Physics (2005).}}

\end{abstract}

\section{Introduction}

We give in this paper a summary of our research program (commenced in
\cite{PRJDissertation}) on the quantum statistical dynamics of relativistic
particles moving in a quantum field. The relativistic particles are not
coupled to each other explicitly but only through their interactions with a
quantum field. Because of this we need to take into consideration both the
effect of each particle on the quantum field and the backreaction of the
quantum field on each particle. A self-consistent treatment of the
particle-field system dynamics is thus necessary. The backreaction from the
field engenders \textit{nonlinear} coupling amongst the particles, and
\textit{non-Markovian} dynamics. We present a new approach to this problem
which highlights the stochastic effects of noise, decoherence, dissipation,
fluctuations, and correlations, and their interconnections.

In \cite{JH0} we presented the basic framework built on the concepts of
quantum open systems \cite{QOS}, the model of quantum Brownian motion (QBM)
\cite{QBM}, and the methodologies of the influence functional \cite{IF},
closed-time-path (in-in) \cite{CTP} coarse-grained effective action
\cite{CGEA,CHMPhysRep}, and world line quantization \cite{WL}. In \cite{JH1}
we applied this framework to spinless relativistic moving particles in a
quantum scalar field, and derived the stochastic equations of motion, known as
the Abraham-Lorentz-Dirac-Langevin (ALDL) equations, for their trajectories
with self-consistent backreaction. The quantum fluctuations of the field are
partially encoded as a stochastic noise in the ALDL equations. The mean
trajectory obtained from taking the stochastic average is governed by
Abraham-Lorentz-Dirac (ALD) equations with modified coefficients whose
time-dependence enforce causality.

Here we use the approach developed in \cite{JH1} to analyze a uniformly
accelerated particle where the trajectory is self-consistently determined by
its interaction with a quantum field (this includes the effects of radiation
reaction and vacuum fluctuations). This important example illustrates the
interconnections between radiation, radiation reaction, dissipation, quantum
fluctuations, and the Unruh effect. Specifically, the uniform acceleration
alters the quantum vacuum which in turn induces thermal-like fluctuations in
the trajectory of a particle. This manifestation of the Unruh effect in terms
of fluctuating trajectories is in contrast to the original formulation and
more common derivations \cite{Unruh76}, where a particle with internal degrees
of freedom (called a detector in the literature \cite{Unruh76,DeWitt75})
moving on a uniformly accelerating prescribed trajectory feels hot at the
Unruh temperature $T_{U}=\hbar a/2\pi ck_{B}$.

Before delving into the technical details, in the remainder of this section we
discuss the outstanding physical issues touched on in this work. In Sec. 2 we
describe the relation of classical radiation, radiation reaction, vacuum
fluctuations and quantum radiation, correcting the misconception of directly
linking vacuum fluctuations with radiation reaction. In Sec. 3 we present the
main structure of our approach based on the open system concept and the
coarse-grained effective action techniques. We show the steps leading to a
derivation of the ALD-Langevin equation for charges moving in a quantum field.
In Sec. 4 we specialize to the case of a particle moving in a background field
which imparts to it uniform acceleration on the average, and derive the Unruh
effect. We discuss how this example can serve as an analog for studying the
stochastic dynamics of black hole event horizons. In Sec. 5 we step back for a
broader perspective and give some discussions of a) the special features of
this new approach and its potential in overcoming the stumbling blocks present
in existing approaches; and b) applications of these results and directions
for further development.

\subsection{Quantum, Stochastic, Semiclassical and Classical}

In this program of investigation we take a microscopic approach and an
{open-systems} perspective, using quantum field theory as the tool to provide
a first-principles derivation of moving particles interacting with a quantum
field. We begin with the closed system of quantum particles and fields. A
\textit{closed system} can be meaningfully partitioned into subsystems if
there exists discrepancies in the scales describing each subsystem, or in
accordance to the physical scales present in their interaction strengths
(energy or time scales) in relation to the probing scale or resolution
accuracy. If we are interested in the details of one such subsystem (e.g.,
particle dynamics) and decide to ignore certain details of the other
subsystems (e.g., details in the quantum field configurations, such as
correlations and phase information) comprising the \textit{environment}, the
distinguished subsystem is rendered an \textit{open system}. The overall
effect of the coarse-grained environment on the open-system can be captured by
the influence functional technique of Feynman and Vernon \cite{IF}, or the
closely related closed-time-path (CTP) effective action method of Schwinger
and Keldysh \cite{CTP} . These are the initial value, or in-in, formulations
suitable for following the time evolution of a system, in contradistinction to
the usual in-out (Schwinger-DeWitt) effective action formulation useful for
S-matrix calculations.

For the model of particle-field interactions under study, coarse-graining the
quantum field yields a nonlocal \textit{coarse-grained effective action
}(CGEA) for the particle motion \cite{CGEA,CHMPhysRep}. An exact expression
may be found in the special case that the particle and field are initially
uncorrelated, and the field state is Gaussian. The CGEA may be used to treat
the fully nonequilibrium quantum dynamics of interacting particles. However,
when higher-order quantum corrections are suppressed by decoherence, the
influence of the quantum fluctuations of the field may be encoded as
stochastic noise, and the CGEA can be transcribed into a \textit{stochastic
effective action} \cite{Banff,Ramsey} describing the stochastic particle
motion. Upon further coarse-graining, taking the stochastic average gives the
semiclassical particle trajectory.

Let us analyze the physics of nonequilibrium processes for such particle-field
systems at separate levels.

\textbf{Classical} level: From the complete quantum microscopic description of
the system (particle) and the environment (field), a classical description is
reached when the system and environment are coarse-grained to an extent that
both the particle and field histories are fully decohered
\cite{GelHar2,conhis}. The fully coarse-grained trajectory for the particle
with backreaction from the fully coarse-grained field then obeys classical
equations of motion, such as the ALD\ equation. The coarse-graining length
scale necessary to achieve full classicality typically far exceeds the length
scale for backreaction pathologies like preacceleration and runaway solutions.

\textbf{Semiclassical} level: This is often defined as a classical system
(particles or detectors) interacting with a quantum environment (quantum
field). It can be obtained from the complete quantum microscopic description
by fully coarse-graining the environment (the quantum field) and finding the
quantum average of the particle trajectory. In the special but important class
when the field enters linearly in the particle-field interaction term, and
without field self-interactions, coarse graining over the quantum field gives
an exact CGEA. The CGEA yields the equations of motion for the system which in
the leading order approximation is the semiclassical limit, followed by
higher-order quantum corrections. In this particular situation, if decoherence
can sufficiently suppress the nonlinear quantum corrections to the
semiclassical equations of motion from the particle, then the semiclassical
limit from the CGEA agrees with the classical limit.

\textbf{Stochastic} level: Going beyond a mean-trajectory (or fully
coarse-grained) description of the particle, a stochastic component in the
particle trajectory appears, as induced by the quantum field fluctuations
manifesting as a classical stochastic noise counterbalancing the quantum
dissipation in the system dynamics. The CGEA may be transcribed into a
stochastic effective action, encoding much of the quantum statistical
information of the field and the state of motion of the system in a noise
correlator for the particle. The stochastic effective action yields classical
stochastic equations of motion for the system which embodies a quantum
dissipation effect (\textit{over and above} the classical radiation reaction)
that balances the quantum fluctuations via a fluctuation-dissipation relation (FDR).

\subsection{Paradoxes and pathologies of backreaction}

\subsubsection{The ALD equations}

The classical theory of moving charges interacting with a classical
electromagnetic (EM) field, when the backreaction of the EM field is included,
has many controversial difficulties \cite{ALD Pathologies}. The generally
accepted classical equation of motion for a charged, spinless point particle
including radiation-reaction is the Abraham-Lorentz-Dirac (ALD) equation
\cite{ALD equation}:
\begin{equation}
\ddot{z}^{\mu}+\tau_{0}(\dot{z}^{\mu}\ddot{z}^{2}+\overset{...}{z}^{\mu
})=(e/mc)\dot{z}_{\nu}F_{ext}^{\mu\nu}(z).
\label{Classical Lorentz-Dirac equation}%
\end{equation}
(See Sec. II for notations and definitions.) The timescale $\tau_{0}=\left(
2e^{2}/3mc^{3}\right)  $ determines the relative importance of the
radiation-reaction term. For electrons, $\tau_{0}\sim10^{-24}$ secs, which is
the time it takes light to cross the electron classical radius $r_{0}%
\sim10^{-15}$m.

The ALD equation is pathological. Because it is a third order differential
equation, it requires the specification of the initial acceleration in
addition to the usual position and velocity required by second order
differential equations. This leads to the existence of runaway solutions.
Physical (e.g. non-runaway) solutions may be enforced by transforming Eq.
(\ref{Classical Lorentz-Dirac equation}) to a second order integral equation
with boundary condition such that the final energy of the particle is finite
and consistent with the total work done on it by external forces. But the
removal of runaway solutions yields acausal solutions that pre-accelerate on
timescales $\tau_{0}.$ This is a source of lingering questions on whether the
classical theory of point particles and fields is causal.

There have been many efforts to understand charged particle radiation-reaction
in the classical and quantum theory. To satisfy the self-consistency and
causality requirements different measures are introduced to the underlying
model. Examples include imposing a high-energy cutoff for the field, an
extended charge distribution, special boundary conditions, particle spin, and
the use of perturbation theory or order reduction techniques \cite{ALD
Solutions}. Previous work closely related to ours include those on quantum
Langevin equations \cite{QLE,FOL}, and the application of the Feynman-Vernon
influence functional to non-relativistic particle coherence \cite{IF: Electron
coherence} and stochastic gravity \cite{StoGraRev} . A major distinction of
our work is the focus on relativistic systems, non-equilibrium processes, and
nonlinear particle-field interaction.

\subsubsection{Coarse-graining and causality}

Consider the situation where a localized particle at rest has some quantum
fluctuations in its position just before an external force is applied at time
$t_{i}$, and where, by chance, the fluctuation is one that seems to anticipate
the force creating the\textit{\ appearance} of preacceleration. Or, consider a
situation where a quantum fluctuation produces the \textit{appearance} of
runaway acceleration for some period of time. While such specific fluctuations
may be highly improbable to realize, they are among the set of fine-grained
solutions in a sum-over-histories formulation of quantum mechanics. These
examples illustrate the observational challenges to distinguishing cause and
effect on a fine-grained quantum scale.

Why is there such a difficulty? The underlying microscopic theory may be
consistent (though for field theory this may not be obvious) and causal, in
the sense that the operator equations of motion are causal, but the quantum
state (for particle or field) is inherently nonlocal. Observables (expectation
values) of the particle motion involve both the operator equations of motion
and the nonlocal quantum state, and thus the question of causality for
trajectories requires more careful treatment. Operationally, we do not observe
fine-grained histories; instead we observe coarse-grained histories where the
coarse-graining scale is set by the measurement resolution in both time and
space. With coarse graining, the quantum fluctuations giving highly
nonclassical trajectories are suppressed, and we therefore expect that a
\textquotedblleft causal and consistent\textquotedblright\ quantum theory
should yield, upon suitable coarse graining, a classical or semiclassical
limit that is pathology free. This is an oversimplified sketch of a more
elaborate theory \cite{HJR} based on the decoherent history approach to
quantum mechanics \cite{conhis}.

\subsection{Uniform acceleration}

\subsubsection{Radiation and radiation reaction}

A particularly interesting class of dynamics from the solutions to the
ALD\ equation is that of a uniformly accelerated particle, defined by $\dot
{z}^{\mu}\dot{z}_{\mu}=a^{2}$. This class of dynamics exemplifies many of the
important issues under discussion. First, it follows from the ALD\ equations
that the classical radiation reaction force vanishes despite the existence of
classical (Lamor) radiation registered as an energy flux at infinity. Hence,
there is no direct balance between radiation and radiation reaction. There is
one existing belief that since radiation is associated with radiation
reaction, the extra work done on the particle against the radiation reaction
force must directly provide the energy that goes into radiation, but this is a
static viewpoint that neglects the full interplay of particle, near field (the
so-called acceleration or Shott field), and far field (the radiation field)
dynamics. Fields are dynamical entities with unusual properties (such as
nonlocality and field correlations) much more complex than just radiation,
which is a far-field definition, and energy can be attributed to a variety of
sources other than radiation. For example, the acceleration field is known to
contain energy and do work. One cannot simply equate work done against
radiation reaction forces with the energy (\textquotedblleft
instantaneously\textquotedblright) radiated into infinity. It would require a
`freezing out' of the near-field's ability to exchange or adjust the form of
its energy consistently in time to be commensurate with the far field behavior
known as radiation.

During periods of uniform acceleration, the energy \textquotedblleft
transfer'\ into radiation comes from acceleration fields, leaving zero
radiation reaction. This is a special situation. During periods of nonuniform
acceleration, there is radiation reaction, and one finds a mixture of field
components; but even then, the energy apportioned to radiation cannot be
instantaneously ascribed to work done against radiation reaction. On matters
related to fields one should look carefully at the energy content locally and
guard against making global statements.

\subsubsection{Quantum radiation, vacuum fluctuations, and the Unruh effect}

It has long been known that a uniformly accelerated detector (UAD) (a detector
is defined as a particle with some internal degree of freedom) coupled to a
quantum field following a prescribed trajectory registers its quantum field
vacuum as a thermal state with temperature $T_{U}=a\hbar/2\pi k_{B}c.$ This is
the Unruh effect \cite{Unruh76}, often cited as an analog of Hawking effect
\cite{Hawking75}. But there are important differences: In the case of a black
hole, real radiation of thermal nature is emitted at the Hawking temperature.
For a uniformly accelerated detector (UAD) there is no emitted radiation
associated with the Unruh effect (see, e.g., \cite{RHA,CapHR} and references
therein) except for a transient period when the charge is coming into
equilibrium with its environment \cite{RHK,CapHJ}. For a uniformly accelerated
\emph{charge} (UAC) there is of course classical radiation, but it is
different from the Unruh radiation, which is a distinctly quantum effect.

We demonstrate below that for a uniformly accelerated charge interacting with
a quantum field, the vacuum fluctuations induce stochastic fluctuations in the
particle \emph{trajectory} with a thermal character. We can refer to this as
the Unruh effect for charges, the role of the \textquotedblleft detector' is
played by the particle motion itself rather than an internal degree of
freedom. In the semiclassical limit (neglecting quantum corrections from the
particle) the \textit{average} emitted radiation is the usual classical
radiation: there is no additional net \textquotedblleft
quantum\textquotedblright\ radiation from the particle. What is new and
interesting in our finding is that there could be radiation associated with
the stochastic component of the particle trajectory. For instance, these could
conceivably generate fluctuations in the emitted radiation, perhaps with a
thermal character, or they could produce other non-classical correlations in
the radiation. This requires further investigation. It has been shown that for
a uniformly accelerated detector on a fixed trajectory the field correlation
is altered around the detector, but as these altered correlations fall off
faster than radiation they can be considered as a vacuum polarization effect
\cite{RHA,Unruh92,MasPar}. Similar vacuum polarization effects are also
expected for a moving charge (contrast this prediction with \cite{GSMP}).
Finally, the radiation emitted by an evaporating (shrinking) black hole under
non-equilibrium conditions is expected to have a non-thermal component
\cite{RHK} over and above the usual thermal (Hawking) part. We are using the
analog with non-uniformly accelerated detectors or charges to investigate this
issue \cite{RHK}.

\section{Semiclassical limit}

\subsection{Coarse-grained effective action}

We begin with the full microscopic theory of a spinless charged particle
interacting with a scalar field. The action for the closed system of particle
plus field is%

\begin{equation}
S\left[  z,\varphi,h\right]  =S\left[  z,h\right]  +S\left[  \varphi\right]
+S\left[  z,\varphi\right]  ,
\end{equation}
where%
\begin{align}
S\left[  z,h\right]   &  =\int d\tau\left(  m_{0}\sqrt{\dot{z}^{\mu}\dot
{z}_{\mu}}+h_{\mu}\left(  \tau\right)  \dot{z}^{\mu}\left(  \tau\right)
\right) \\
S\left[  \varphi\right]   &  =\int d^{4}x\,\frac{1}{2}\left(  \partial_{\mu
}\varphi\right)  ^{2}\\
S\left[  z,\varphi\right]   &  =\int d^{4}x\,j\left(  x;z\right)  \left(
\varphi\left(  x\right)  +\varphi_{ext}\left(  x\right)  \right)  .
\end{align}
The spacetime worldline coordinates of the particle are $z^{\mu}\left(
\tau\right)  ,$ $\varphi\left(  x\right)  $ is the scalar field$,$ and
$\varphi_{ext}\left(  x\right)  $ is an external field. The functions $h_{\mu
}\left(  \tau\right)  $ are auxiliary sources introduced to generate
correlation functions, after which they are set to zero. The field couples to
the particle through the reparametrization invariant scalar current%
\begin{equation}
j\left(  x;z\right)  =e\int d\tau\sqrt{\dot{z}^{\mu}\dot{z}_{\mu}}%
\delta\left(  x-z\left(  \tau\right)  \right)  .
\end{equation}
The action $S\left[  z,\varphi,h\right]  $ is manifestly invariant under
reparametrizations $\tau\rightarrow\tau^{\prime}=\tau+\varepsilon\left(
\tau\right)  .$ The particle mass is $m_{0}$ and charge is $e.$ We use units
where $\hbar=c=1$, Greek indices $\mu,\nu,...=\left(  0,1,2,3\right)  ,$ and
$g_{\mu\nu}=\left(  1,-1,-1,-1\right)  .$

Consider an initial density matrix for the particle plus field at time $t_{i}$
that has the factorized form
\begin{equation}
\hat{\rho}_{i}=\hat{\rho}_{z}\left(  t_{i}\right)  \otimes\hat{\rho}_{\varphi
}\left(  t_{i}\right)  , \label{initial density matrix}%
\end{equation}
where $\hat{\rho}_{z}\left(  t_{i}\right)  $ is the initial density matrix for
the particle subsystem $\left(  z\right)  $ and $\hat{\rho}_{\varphi}\left(
t_{i}\right)  $ is the initial density matrix for the field subsystem $\left(
\varphi\right)  $. We note at the start that strictly speaking such factorized
states are unphysical because they represent a completely decoupled particle
and field. More general initial states could be considered, but treating a
fully physical state remains an important challenge \cite{RomeroPaz96}. For
$\hat{\rho}_{\varphi}\left(  t_{i}\right)  $ we assume that the initial field
state is Gaussian (this includes thermal, squeezed, and coherent states, and
therefore provides us with a fairly rich set of interesting and physical
examples.) A relativistic covariant particle state $\left\vert z\right\rangle
\equiv\hat{\psi}^{\left(  +\right)  \dagger}\left(  z\right)  \left\vert
0\right\rangle ,$ where $\hat{\psi}^{\left(  +\right)  \dagger}\left(
z\right)  $ is the positive frequency operator for creating a field excitation
centered at $z,$ has a localization size (in terms of the charge-density
expectation value) of a Compton wavelength $\lambda_{c}=h/mc,$ imparting a
natural \emph{minimum} coarse-graining scale. Typical particle wavepackets
have much larger (characteristic) de Broglie wavelengths, and thus require
considerable further averaging of fine-grained trajectories to achieve a
coarse-grained decoherent history. We assume that the initial particle state
is a wavepacket localized around $\mathbf{z}_{i}$ with width $\Lambda
\gg\lambda_{c}.$ We will see that the degree of coarse graining does (weakly)
effect the semiclassical trajectory via higher derivative terms that can
generally be neglected. (Note that both $\Lambda$ and $\lambda_{c}$ are far
larger that the classical radius of a charged particle $r_{0},$ the scale at
which pathological backreaction behavior manifests).

For the closed system, the density matrix at a later time $t_{f}$ is%
\begin{equation}
\hat{\rho}\left(  t_{f}\right)  =\hat{U}\left(  t_{f},t_{i}\right)  \hat{\rho
}_{i}\hat{U}\left(  t_{f},t_{i}\right)  ^{\dagger},
\end{equation}
where $\hat{U}$ is the unitary evolution operator for the fully interacting
particle plus field system. The matrix elements of $\hat{U}$ can be given a
sum over histories representation:%
\begin{align}
K_{h}\left(  \varphi_{f},\mathbf{z}_{f},t_{f};\varphi_{i},\mathbf{z}_{i}%
,t_{i}\right)   &  \equiv\left\langle \varphi_{f},\mathbf{z}_{f}\right\vert
\hat{U}_{h}\left(  t_{f},t_{i}\right)  \left\vert \varphi_{i},\mathbf{z}%
_{i}\right\rangle \nonumber\\
&  =\int_{\varphi_{i},\mathbf{z}_{i}}^{\varphi_{f},\mathbf{z}_{f}}D\varphi
Dz\exp\left\{  iS\left[  z,\varphi,h\right]  \right\}  ,
\end{align}
where the functional measures are%
\begin{align}
\int_{\varphi_{i}}^{\varphi_{f}}D\varphi &  =\int\prod_{x}d\varphi\left(
x\right)  ,\\
\int_{z_{i}}^{z_{f}}Dz  &  =\int\prod_{\mu,\tau}dz^{\mu}\left(  \tau\right)  ,
\end{align}
subject to appropriate boundary conditions that are discussed in \cite{JH1}.
The in-in or closed-time-path (CTP) generating functional is given by
\begin{align}
Z_{\text{in-in}}\left[  h,h^{\prime}\right]   &  =\text{Tr}_{\varphi
,z}\text{\ }\left[  \hat{U}_{h}\left(  t_{f},t_{i}\right)  \hat{\rho}_{i}%
\hat{U}_{h^{\prime}}\left(  t_{f},t_{i}\right)  ^{\dagger}\right] \\
&  =\int d\mathbf{z}_{f}d\varphi_{f}d\varphi_{i}d\varphi_{i}^{\prime
}d\mathbf{z}_{i}d\mathbf{z}_{i}^{\prime}\text{\ }\nonumber\\
&  \times K_{h}\left(  \varphi_{f},\mathbf{z}_{f},t_{f};\varphi_{i}%
,\mathbf{z}_{i},t_{i}\right)  K_{h^{\prime}}^{\ast}\left(  \varphi
_{f},\mathbf{z}_{f},t_{f};\varphi_{i}^{\prime},\mathbf{z}_{i}^{\prime}%
,t_{i}\right) \nonumber\\
&  \times\rho\left(  \varphi_{i},\mathbf{z}_{i};\varphi_{i}^{\prime
},\mathbf{z}_{i}^{\prime};t_{i}\right)  .\nonumber
\end{align}
This generating functional is a tool for computing the average particle
trajectory and its fluctuations (i.e. the correlation functions for the
particle worldline coordinates) including the effects of backreaction from the
coarse-grained field self-consistently.

Using the assumed initially factorized state and the sum-over-histories
representation, the generating functional may be written as%
\begin{equation}
Z_{\text{in-in}}\left[  h,h^{\prime}\right]  =\int DzDz^{\prime}e^{i\left\{
S\left[  z,h\right]  -S\left[  z^{\prime},h^{\prime}\right]  \right\}
}F\left[  z,z^{\prime}\right]  ,
\end{equation}
where%
\begin{align}
F\left[  z,z^{\prime}\right]   &  =\int D\varphi D\varphi^{\prime}e^{i\left\{
S\left[  \varphi\right]  +S\left[  z,\varphi\right]  -S\left[  \varphi
^{\prime}\right]  -S\left[  z^{\prime},\varphi^{\prime}\right]  \right\}
}\rho_{i}\left(  \varphi_{i},\varphi_{i}^{\prime},t_{i}\right) \nonumber\\
&  \equiv\exp\left\{  iS_{IF}\left[  z,z^{\prime}\right]  \right\}
\end{align}
is the Feynman-Vernon influence functional, and $S_{IF}\left[  z,z^{\prime
}\right]  $ is the influence action. From the definition of the in-in
generating functional it follows that%
\begin{equation}
\left(  \frac{1}{i}\right)  ^{n}\left.  \frac{\delta^{n}Z_{\text{in-in}%
}\left[  h,h^{\prime}\right]  }{\delta h_{\mu}\left(  \tau_{n}\right)
...\delta h_{\mu}\left(  \tau_{1}\right)  }\right\vert _{h,h^{\prime}%
=0}=\text{Tr}_{\varphi,z}\langle\hat{z}^{\mu}\left(  \tau_{n}\right)
...\hat{z}^{\nu}\left(  \tau_{1}\right)  \hat{\rho}_{i}\rangle.
\end{equation}
Hence, we see that $Z_{\text{in-in}}$ is indeed a generating function for the
expectation values of the worldline coordinates at an arbitrary proper time
$\tau_{i}.$

Evaluating the full generating functional is virtually impossible. The
interaction term $S\left[  z,\varphi\right]  ,$ while linear in $\varphi,$ is
highly non-linear in the worldline coordinate $z.$ However, if the initial
state of the field $\rho_{\varphi}\left(  t_{i}\right)  $ is Gaussian, the
influence functional involves purely Gaussian integrands in the variables
$\varphi$, and thus can be evaluated exactly, given the well-known result (see
e.g., \cite{RHK})
\begin{equation}
S_{IF}=\int d^{4}xd^{4}x^{\prime}\left[  j^{\left(  -\right)  }\left(
x\right)  G_{R}\left(  x,x^{\prime}\right)  j^{\left(  +\right)  }\left(
x^{\prime}\right)  +\frac{i}{4}\jmath^{\left(  -\right)  }\left(  x\right)
G_{H}\left(  x,x^{\prime}\right)  j^{\left(  -\right)  }\left(  x^{\prime
}\right)  \right]  ,
\end{equation}
where
\begin{align}
G_{R}\left(  x,x^{\prime}\right)   &  =Tr_{\hat{\varphi}}\left(  i\left[
\hat{\varphi}\left(  x\right)  ,\hat{\varphi}\left(  x^{\prime}\right)
\right]  \hat{\rho}_{A}\left(  t_{i}\right)  \right)  \theta\left(
t,t^{\prime}\right)  ,\\
G_{H}\left(  x,x^{\prime}\right)   &  =Tr_{\hat{\varphi}}\left(  \left\{
\hat{\varphi}\left(  x\right)  ,\hat{\varphi}\left(  x^{\prime}\right)
\right\}  \hat{\rho}_{\hat{\varphi}}\left(  t_{i}\right)  \right)
\end{align}
are the retarded and Hadamard Green's functions, respectively. The brackets
$\left[  ,\right]  $ and $\left\{  ,\right\}  $ denote commutators and
anticommutators. The sum and difference currents $j^{\left(  \pm\right)  }$
are defined by
\begin{align}
j^{\left(  -\right)  }\left(  x;z,z^{\prime}\right)   &  =\left(  j\left(
x\right)  -j^{\prime}\left(  x\right)  \right) \\
j^{\left(  +\right)  }\left(  x;z,z^{\prime}\right)   &  =\left(  j\left(
x\right)  +j^{\prime}\left(  x\right)  \right)  /2.
\end{align}
We also define%
\begin{align}
z^{\left(  -\right)  }\left(  \tau\right)   &  =\left(  z\left(  \tau\right)
-z^{\prime}\left(  \tau\right)  \right) \\
z^{\left(  +\right)  }\left(  \tau\right)   &  =\left(  z\left(  \tau\right)
+z^{\prime}\left(  \tau\right)  \right)  /2,
\end{align}
and%
\begin{align}
h^{\left(  -\right)  }\left(  \tau\right)   &  =\left(  h\left(  \tau\right)
-h^{\prime}\left(  \tau\right)  \right) \\
h^{\left(  +\right)  }\left(  \tau\right)   &  =\left(  h\left(  \tau\right)
+h^{\prime}\left(  \tau\right)  \right)  /2.
\end{align}
The in-in generating functional can be compactly expressed as%

\begin{equation}
Z_{\text{in-in}}\left[  h^{\left(  \pm\right)  }\right]  =\int Dz^{\left(
+\right)  }Dz^{\left(  -\right)  }e^{iS_{CGEA}\left[  z^{\left(  \pm\right)
},h^{\left(  \pm\right)  }\right]  }, \label{Full generating functional}%
\end{equation}
where
\begin{equation}
S_{CGEA}\left[  z^{\left(  \pm\right)  },h^{\left(  \pm\right)  }\right]
=S_{z}\left[  z,h\right]  -S_{z}\left[  z^{\prime},h^{\prime}\right]
+S_{IF}\left[  z^{\left(  \pm\right)  }\right]
\end{equation}
is the in-in or CTP coarse-grained effective action (CTPCGEA). We emphasize
that $S_{CGEA}$ encapsules the full effects of the coarse-grained quantum
field, and $Z_{\text{in-in}}\left[  h^{\left(  \pm\right)  }\right]  $
generates, in principle, the full information about the quantum dynamics of
the particle. The semiclassical description of the particle dynamics is given
by the particle trajectory's expectation value $\bar{z}^{\mu}=\left\langle
\hat{z}^{\mu}\left(  \tau\right)  \right\rangle ,$ neglecting particle quantum
corrections, obtained from the equations of motion%
\begin{equation}
\left.  \frac{\delta S_{CGEA}}{\delta z_{\mu}^{\left(  -\right)  }\left(
\tau\right)  }\right\vert _{z^{\left(  -\right)  }=0,z^{\left(  +\right)
}\equiv\bar{z}}=0. \label{Semiclassical eq from SCEA}%
\end{equation}
Upon setting $z^{\left(  -\right)  }=0$ the imaginary part of $S_{CGEA},$
which involves the Hadamard function, gives vanishing contribution.
Consequently, the semiclassical equations of motion depends only on the
retarded green's function and are therefore real and causal.

\subsection{Modified ALD equation with time-dependent coefficients}

The semiclassical trajectory is a solution of the equation derived from Eq.
(\ref{Semiclassical eq from SCEA}) \cite{Lin}
\begin{equation}
m_{0}\ddot{z}_{\mu}=\frac{e^{2}}{c^{3}}\int d\tau^{\prime}\vec{w}_{\mu}%
G^{R}\left(  z\left(  \tau\right)  ,z\left(  \tau^{\prime}\right)  \right)
+e\vec{w}_{\mu}\varphi_{ext}\left(  z\right)  ,
\end{equation}
where%
\begin{equation}
\vec{w}_{\mu}\equiv\dot{z}^{\nu}\dot{z}_{[\nu}\partial_{\mu]}-\ddot{z}_{\mu}.
\end{equation}
The external potential $\varphi_{ext}$ provides an external force $F_{\mu
}^{ext}=e\vec{w}_{\mu}\varphi_{ext}$ with two main components: a gradient
force $e\dot{z}^{\nu}\dot{z}_{[\nu}\partial_{\mu]}\varphi_{ext}=e\left(
g_{\mu\nu}-\dot{z}_{\mu}\dot{z}_{\nu}\right)  \partial^{\nu}\varphi_{ext},$
and an effective contribution to the particle mass, $e\varphi_{ext}\ddot
{z}_{\mu}.$ Note that $\dot{z}^{\mu}\vec{w}_{\mu}=\dot{z}^{\mu}\ddot{z}_{\mu
}=0$ identically, and therefore $\dot{z}^{\mu}F_{\mu}^{ext}=0,$ as required of
a relativistic force that preserves the mass-shell constraint $\dot{z}^{2}=1.$
The scalar field force $F_{\mu}^{ext} $ is analogous to the electromagnetic
force $F_{\mu}^{EM}=\dot{z}^{\nu}F_{\mu\nu}$ which satisfies $\dot{z}^{\mu
}F_{\mu}^{EM}=0.$

The retarded Green's function is singular when $\tau=\tau^{\prime},$ and so
requires regularization. We choose%
\begin{equation}
G_{R}^{\Omega}\left(  \sigma\right)  =\frac{\Omega J_{1}\left(  \Omega
\sigma\right)  }{4\pi\sigma},
\end{equation}
where $J_{1}$ is a first order Bessel function and $\sigma^{2}=\left(
z-z^{\prime}\right)  ^{\mu}\left(  z-z^{\prime}\right)  _{\mu}.$ In the limit
$\Omega\rightarrow\infty:G_{R}^{\Omega}\rightarrow\delta\left(  \sigma
^{2}\right)  .$ Consistency requires that the Hadamard Green's function is
regulated with the same effective cutoff: $G_{H}\rightarrow G_{H}^{\Omega}.$
We assume that the high-frequency mode cutoff $\Omega$ comes from the limit of
the measurement resolution or the preparation scale for the initial particle
state, so that $\Omega\sim1/\Lambda.$ It is reasonable to assume (though this
argument needs to be made rigorous) that on finer-grained scales the
correlations of the particle and field are undisturbed by the motion of the
particle. This suggests viewing the initial factorized state as representing a
particle that, through the initial state preparation process at time $t_{i}$,
is uncorrelated only with the \emph{longer} wavelength modes of field, while
it remains fully dressed by the (undisturbed) short wavelength modes. The
regulated Green's function is therefore a model for the response of the long
wavelength field modes to the particle motion, rather than a short wavelength
(high energy) modification of the underlying quantum field theory.

Using our explicit choice for $G_{\Omega}^{R}\left(  \sigma\right)  $, the
resulting equations of motion are%
\begin{equation}
m\left(  r\right)  \ddot{z}_{\mu}\left(  \tau\right)  =e\vec{w}_{\mu}%
\varphi_{ext}+\frac{\tau_{0}}{2}f_{\Omega}^{\left(  0\right)  }(r)\left(
\dot{z}_{\mu}\ddot{z}^{2}+\dddot{z}_{\mu}\right)  +\frac{\tau_{0}}{2}%
\sum_{n=1}^{\infty}\frac{f_{\Omega}^{\left(  n\right)  }(r)}{\Omega^{n}}%
u_{\mu}^{\left(  n\right)  }, \label{Full semiclassical limit}%
\end{equation}
where $r$ $=\tau-\tau_{i}$ is the elapsed proper-time from the initial time
$\tau_{i}=\tau_{i}\left(  t_{i}\right)  $ [Recall that $\tau_{0}=\left(
2e^{2}/3mc^{3}\right)  $]. The time-dependent radiation reaction term
proportional to $\Omega\ddot{z}_{\mu}$ has been absorbed into a time-dependent
mass $m\left(  r\right)  .$ At $r=0$, corresponding to the initial time when
the field and particle are uncorrelated, $m\left(  0\right)  =m_{0}$ ($m_{0}$
is the bare mass which includes the renormalization from the short wavelength
field modes above the cutoff $\Omega$). The field dresses the particle,
changing its effective mass, which reaches an asymptotic value $m\equiv
m\left(  \infty\right)  =m_{0}-e^{2}\Omega/8\pi c^{3}$ on the cutoff timescale
$1/\Omega.$ It is interesting to note that this shift is $-1/2$ times the
shift that results for an electromagnetic field. Since the EM field may be
viewed as two independent scalar fields (one for each polarization), the
factor of $1/2$ for a scalar field is intuitively reasonable. A negative
rather than positive mass shift comes about due to differences between how
scalar and EM fields couple to a particle.

Radiation reaction is given by the last two terms on the right side (RHS) of
Eq. (\ref{Full semiclassical limit}). The second term has the usual third
derivative form for ALD\ radiation reaction, except for the time-dependent
coefficient $f_{\Omega}^{\left(  0\right)  }\left(  r\right)  .$ This function
depends on the exact choice of the regulated retarded Green's function, but it
has the general property that $f_{\Omega}^{\left(  0\right)  }\left(
0\right)  =0$ and $f_{\Omega}^{\left(  0\right)  }\left(  \infty\right)  =1,$
with the timescale for the transition from 0 to 1 being $1/\Omega.$ The last
term on the RHS of Eq. (\ref{Full semiclassical limit}) represent a set of
higher derivative contributions to radiation reaction. These are suppressed at
low energies by factors of $1/\Omega^{n}$. The functions $f_{\Omega}^{\left(
n\right)  }(r)$ vanish when $r=0,$ and go to constant values on a timescale
$1/\Omega.$ At the initial time $r=0$ the radiation reaction terms, including
the higher derivative contributions, vanish, and this makes the solutions to
Eq. (\ref{Full semiclassical limit}) unique (requiring only the usual position
and velocity initial data), causal (no preacceleration), and runaway free.

The functions $u_{\mu}^{\left(  n\right)  }$ involve fourth and higher
derivative contributions:%
\begin{equation}
u_{\mu}^{\left(  n\right)  }=u_{\mu}^{\left(  n\right)  }\left(  \frac
{d^{n+3}z}{d\tau^{n+3}},...,\frac{dz}{d\tau}\right)  .
\end{equation}
The dependence of the equations of motion in Eq.
(\ref{Full semiclassical limit}) on $\Omega$ and the higher derivative terms
reflects the degree to which the particle subsystem (trajectory) is
distinguished from the full system (particle and field) by coarse graining.
The larger $\Omega$ the more point-like the particle. For sufficiently
point-like particles (finer grained trajectories) and longer time scales
($r\gg1/\Omega$), the higher derivative terms are strongly suppressed, and the
dominant contribution to the equations of motion are independent of $\Omega,$ giving%

\begin{equation}
m\ddot{z}_{\mu}\left(  \tau\right)  =e\vec{w}_{\mu}\varphi_{ext}\left(
z\right)  +\frac{\tau_{0}}{2}\left(  \dot{z}_{\mu}\ddot{z}^{2}+\dddot{z}_{\mu
}\right)  . \label{Scalar ALD equation}%
\end{equation}
Thus, at low energies and long times, the radiation reaction force is of the
usual ALD\ form. (Note that the coefficient $\tau_{0}/2$ is $1/2$ of the value
for radiation reaction in the EM field. Again viewing the EM\ field as
equivalent to two scalar fields, this result is natural.) Since in this limit
$\tau_{0}$ is a very small parameter, in practice we can treat Eq.
(\ref{Scalar ALD equation}) perturbatively. The lowest order solution is found
by neglecting the radiation reaction term so that the trajectory is determined
by the background field $\varphi_{ext},$ giving%
\begin{equation}
\ddot{z}_{\mu}=e\left(  m+e\varphi_{ext}\left(  z\right)  \right)  ^{-1}%
\dot{z}^{\nu}\dot{z}_{[\nu}\partial_{\mu]}\varphi_{ext}\left(  z\right)  .
\label{iterative approx to z''}%
\end{equation}
Taking one proper-time derivative gives%
\begin{equation}
\dddot{z}_{\mu}=\frac{e}{m}\frac{d}{d\tau}\left(  \left(  m+e\varphi
_{ext}\left(  z\right)  \right)  ^{-1}\dot{z}^{\nu}\dot{z}_{[\nu}\partial
_{\mu]}\varphi_{ext}\left(  z\right)  \right)  .
\label{iterative approx to z'''}%
\end{equation}
These two expression for $\ddot{z}$ and $\dddot{z}$ can be substituted into
the radiation reaction term in Eq. (\ref{Scalar ALD equation}), yielding an
order $e^{4}$ equation for the radiation reaction force containing only first
derivatives $\dot{z}$. This procedure can be iterated to produce a
runaway-free and causal perturbative approximation to any order in $e,$ but
clearly this expansion which contains only first derivatives $\dot{z}$ can not
converge to the third-derivative ALD equation. To fully understand causality
the non-Markovian short-time behavior must be examined, with
Eq.(\ref{Full semiclassical limit}) being an example of how non-Markovian
effects can enforce causality.

We have analyzed the special case when the particle and field are initially
uncorrelated below some coarse grained preparation scale $\Omega.$ The
nonequilibrium time-dependence of the radiation reaction force reflects the
fact that in any nonequilibrium quantum setting it takes time for the
particle's self-field to adjust to changes in its motion. This lag preserves
causality unless one takes the cutoff to infinity (finest grained histories):
but this would imply that the high-energy structure of the fundamental theory
is critical to the equations of motion even at low energy. This, while in
principle possible, is in contradiction with our experience that realistic
physical conditions are described by low-energy effective theories. Our
analysis is a first step in demonstrating the full causality of the
coarse-grained semiclassical limit of particle motion in quantum field theory.
Next steps include considering more general initial conditions, rigorously
deriving a coarse-grained (regulated) Green's function from a physical
coarse-graining mechanism, and analyzing the effects of higher-order quantum corrections.

\section{Stochastic limit}

\subsection{Stochastic effective action}

We now go beyond the average trajectory and include the trajectory
fluctuations induced by the quantum fluctuations of the field. To do this, we
return to the full quantum generating function, Eq.
(\ref{Full generating functional}). In practice, to make use of the in-in
generating functional the non-linear $z^{\left(  \pm\right)  }$ path integrals
must be evaluated using some approximation. To find quantum corrections to the
semiclassical solution, the loop expansion is often invoked.
We describe here a different approach that transforms $S_{CGEA}$ into a
stochastic effective action.

Noting that $G_{H}^{\Omega}\left(  x,x^{\prime}\right)  $ is real and
symmetric, and therefore has positive eigenvalues, it follows that
\begin{equation}
\operatorname{Im}\left(  S_{IF}\left[  z^{\left(  \pm\right)  }\right]
\right)  =\frac{1}{4}\int\int dxdx^{\prime}\jmath^{\left(  -\right)  }\left(
x\right)  G_{H}^{\Omega}\left(  x,x^{\prime}\right)  j^{\left(  -\right)
}\left(  x^{\prime}\right)  \geq0,
\end{equation}
and%
\begin{equation}
\left\vert F\left[  z^{\left(  \pm\right)  }\right]  \right\vert =\left\vert
e^{iS_{CGEA}}\right\vert =e^{-Im\left(  S_{IF}\left[  z^{\left(  \pm\right)
}\right]  \right)  }<1.
\end{equation}
The influence functional is suppressed for large $j^{\left(  -\right)  }$,
which reflects the decoherence associated with the radiation produced by the
particle current. Consistency with the cutoff $\Omega$ requires that the
current $j^{\left(  -\right)  }$ does not produce radiation with frequency
greater than $\Omega.$ The contribution to $S_{CGEA}$ from the imaginary part
of the influence functional may be re-written as%
\begin{equation}
\left\vert e^{iS_{CGEA}}\right\vert =\int D\tilde{\varphi}P\left[
\tilde{\varphi}\right]  e^{i\int j^{\left(  -\right)  }\left(  x\right)
\tilde{\varphi}\left(  x\right)  dx},
\end{equation}
where%
\begin{equation}
P\left[  \tilde{\varphi}\right]  =Ne^{-\int dxdx^{\prime}\tilde{\varphi
}\left(  x\right)  G_{H}^{\Omega}\left(  x,x^{\prime}\right)  ^{-1}%
\tilde{\varphi}\left(  x^{\prime}\right)  }%
\end{equation}
is a normalizable probability distribution for a \textit{stochastic field}
$\tilde{\varphi}\left(  x\right)  .$ The two-point function for the stochastic
field is%
\begin{equation}
\left\langle \left\{  \tilde{\varphi}\left(  x\right)  ,\tilde{\varphi}\left(
x^{\prime}\right)  \right\}  \right\rangle _{\text{stoch}}=G_{H}^{\Omega
}\left(  x,x^{\prime}\right)  =\left\langle \left\{  \hat{\varphi}\left(
x\right)  ,\hat{\varphi}\left(  x^{\prime}\right)  \right\}  \right\rangle
_{\text{quantum}},
\end{equation}
showing that $\tilde{\varphi}$ encodes the same information as the quantum
field anti-correlation function, with noise (or correlations) above the cutoff
assumed to be coarse grained and averaged out.

The in-in generating functional may be re-expressed as%
\begin{equation}
Z_{\text{in-in}}\left[  h^{\left(  \pm\right)  }\right]  =\int Dz^{\left(
+\right)  }Dz^{\left(  -\right)  }\int D\tilde{\varphi}P\left[  \tilde
{\varphi}\right]  e^{iS_{stoch}\left[  z^{\left(  \pm\right)  };\tilde
{\varphi}\right]  },
\end{equation}
where%
\begin{align}
S_{stoch}\left[  z^{\left(  \pm\right)  };\tilde{\varphi}\right]   &
=\operatorname{Re}\left\{  S_{CGEA}\left[  z^{\left(  \pm\right)  }\right]
\right\}  +\int j^{\left(  -\right)  }\left(  x\right)  \tilde{\varphi}\left(
x\right)  dx\label{Stochastic effective action}\\
&  =S_{z}\left[  z,h\right]  -S_{z}\left[  z^{\prime},h^{\prime}\right]  +\int
dxj^{\left(  -\right)  }\left(  x\right)  \left(  \tilde{\varphi}\left(
x\right)  +\int dx^{\prime}G_{R}^{\Omega}\left(  x,x^{\prime}\right)
j^{\left(  +\right)  }\left(  x^{\prime}\right)  \right) \nonumber
\end{align}
is the stochastic effective action.

The generating functional now involves both a sum over trajectories and a sum
over stochastic field configurations. If we neglect higher order quantum
corrections and take $S_{stoch}\left[  z^{\left(  \pm\right)  };\tilde
{\varphi}\right]  \simeq S_{stoch}\left[  z_{0}^{\pm};\tilde{\varphi}\right]
,$ where $z_{0}^{\left(  \pm\right)  }$ are the classical (extremal
solutions), then we obtain a stochastic generating functional:%
\begin{equation}
Z_{\text{in-in}}^{stoch}\left[  h^{\left(  \pm\right)  }\right]  =\int
D\tilde{\varphi}P\left[  \tilde{\varphi}\right]  e^{iS_{stoch}\left[
z_{0}^{\left(  \pm\right)  };\tilde{\varphi}\right]  }.
\end{equation}
This describes the stochastic regime, where higher order quantum effects are
washed out by decoherence but some measure of quantum field induced
stochasticity remains. We emphasize that this is not a stochastic theory of
scalar field QED where stochasticity is introduced in an \textit{ad hoc }
manner.
Rather, we have started with a microscopic quantum theory and shown the
conditions (namely strong decoherence that suppresses higher order
corrections) whereby its behavior can be captured by its stochastic effects.

\subsection{ALD-Langevin (ALDL) equations}

The stochastic effective action generates nonlinear stochastic equations of
motion, defined by%
\begin{equation}
\left.  \frac{\delta S_{stoch}}{\delta z_{\mu}^{\left(  -\right)  }\left(
\tau\right)  }\right\vert _{z^{\left(  -\right)  }=0,z^{\left(  +\right)
}=\tilde{z}}=0\text{,} \label{Equations of motion from stochastic action}%
\end{equation}
where $z$ is now taken to be a stochastic variable. Using Eqs.
(\ref{Stochastic effective action}) and
(\ref{Equations of motion from stochastic action}), we find that%

\begin{equation}
m_{0}\ddot{z}_{\mu}=e\vec{w}_{\mu}\left(  \varphi_{ext}+\tilde{\varphi}\left(
z\right)  +\frac{e}{2}\int^{\tau}d\tau^{\prime}G_{R}^{\Omega}\left(  z\left(
\tau\right)  ,z\left(  \tau^{\prime}\right)  \right)  \right)  .
\end{equation}
Or, using the regulated Green's function and dropping higher derivative terms,
the nonlinear stochastic equations of motion take the form (in the late-time
limit $\tau\gg1/\Omega$)%
\begin{equation}
m\ddot{z}_{\mu}=e\vec{w}_{\mu}\left[  \varphi_{ext}\left(  z\right)
+\tilde{\varphi}\left(  z\right)  \right]  +\frac{\tau_{0}}{2}\left(  \dot
{z}_{\mu}\ddot{z}^{2}+\dddot{z}_{\mu}\right)  . \label{Nonlinear ALDL}%
\end{equation}
The perturbative procedure illustrated in Eqs. (\ref{iterative approx to z''})
and (\ref{iterative approx to z'''}) can be similarly be applied to Eq.
(\ref{Nonlinear ALDL}) to obtain first-derivative equations of motion.

The stochastic field provides a stochastic force $\eta_{\mu}\left(
\tau\right)  =e\vec{w}_{\mu}\tilde{\varphi}\left(  z\right)  $ which, just
like the force from $\varphi_{ext},$ satisfies $\dot{z}^{\mu}\eta_{\mu}=0,$
and hence also preserves the mass-shell constraint. Note that the stochastic
field generates an effective stochastic mass term, $e\tilde{\varphi}\left(
z\right)  \ddot{z}_{\mu},$ in parallel with the effective mass term generated
by $\varphi_{ext}.$ In the short-time limit these equations are (like the
semiclassical equations) modified by nonequilibrium (time-dependent) effects.

The noise $\eta_{\mu}\left(  \tau\right)  $ enters the Langevin equation in a
highly nonlinear way, and its correlator is conditional on the particle
history, making the equation very complicated. Since the notion of a
semiclassical trajectory already requires substantial coarse-graining, which
averages out short time and higher order quantum effects, we make the analog
of the background field separation by expanding the equations around the
semiclassical trajectory $\bar{z}_{\mu}\left(  \tau\right)  $ to obtain
linearized Langevin equations for the trajectory fluctuations $y^{\mu}\left(
\tau\right)  \equiv z^{\mu}\left(  \tau\right)  -\bar{z}^{\mu}\left(
\tau\right)  $. This gives%
\begin{equation}
m\ddot{y}_{\mu}=f_{\mu}^{ext}\left(  y\right)  +\int^{\tau}d\tau^{\prime
}D_{\mu\nu}\left(  \tau,\tau^{\prime}\right)  y^{\nu}\left(  \tau^{\prime
}\right)  +\eta_{\mu}. \label{Linearized ALD Langevin}%
\end{equation}
The linearized noise $\eta_{\mu}\left(  \tau\right)  \equiv e\vec{w}_{\mu
}\tilde{\varphi}\left(  \bar{z}\right)  $ has vanishing mean, $\left\langle
\eta^{\mu}\left(  \tau\right)  \right\rangle =0,$ and a two-point correlator%
\begin{equation}
\left\langle \left\{  \eta^{\mu}\left(  \tau\right)  ,\eta^{\nu}\left(
\tau^{\prime}\right)  \right\}  \right\rangle \equiv e^{2}\vec{w}^{\mu}\left(
\bar{z}\left(  \tau\right)  \right)  \vec{w}^{\nu}\left(  \bar{z}\left(
\tau^{\prime}\right)  \right)  G_{H}^{\Omega}\left(  \bar{z}\left(
\tau\right)  ,\bar{z}\left(  \tau^{\prime}\right)  \right)
\label{Noise correlator}%
\end{equation}
that is independent of the particle fluctuations $y$, but is dependent upon
the self-consistently determined semiclassical trajectory $\bar{z}_{\mu
}\left(  \tau\right)  .$ The operator $\vec{w}_{\mu}$ is evaluated in terms of
the solution $\bar{z}.$ The linear force from the external field is given by
\begin{equation}
f_{\mu}^{ext}\left(  y\right)  \equiv\int^{\tau}d\tau^{\prime}\frac{\delta
}{\delta z^{\nu}\left(  \tau^{\prime}\right)  }\left\{  e\vec{w}_{\mu}%
\varphi_{ext}\left(  \bar{z}\right)  \right\}  y^{\nu}\left(  \tau^{\prime
}\right)  .
\end{equation}
The second term on the RHS of Eq. (\ref{Linearized ALD Langevin}) represents
the radiation reaction force, where
\begin{equation}
D_{\mu\nu}\left(  \tau,\tau^{\prime}\right)  \equiv\frac{\tau_{0}}{2}%
\frac{\delta}{\delta z^{\nu}\left(  \tau^{\prime}\right)  }\left\{  \dot
{z}_{\mu}\ddot{z}^{2}+\dddot{z}_{\mu}\right\}
\end{equation}
is the dissipation kernel for linearized fluctuations of the trajectory around
its average (a first-derivative perturbative form of $D_{\mu\nu}$ can be
obtained using Eqs. (\ref{iterative approx to z''}) and
(\ref{iterative approx to z'''}) to remove the $\ddot{z}$ and $\dddot{z}$
factors). The kernel $D_{\mu\nu}$ represents what we have described as quantum
(or stochastic) dissipation: the backreaction in response to the quantum field
induced fluctuations. While there is no direct link between the classical
(ALD) radiation reaction force and quantum fluctuations, there is a
relationship between the \emph{quantum dissipation} and quantum flucutations
\cite{CapHR,CapHJ}.

\section{Uniformly accelerated charges}

In a weak electric field, the vacuum of an uniformly accelerated charge (UAC)
assumes thermal attributes (the Unruh effect) and additional fluctuations are
induced in the particle motion. In a stronger field, charged particle
pair-creation becomes important. These are two distinct types of particle
production processes. In the semiclassical-stochastic limit (in which we have
been working) one does not have charged-particle creation effects. In the
analogous situation of semiclassical gravity and Hawking radiation this
entails the neglect of quantum gravitational effects. For sufficiently high
intensity fields, charged particle pair creation can become a significant
process, and the creation of both photons and charges must be treated as a
fully quantum problem. The gravitation analogy would entail calculations based
on some theory of quantum gravity. There, the emergence of classical spacetime
from a theory of quantum gravity is an issue of fundamental importance. Prior
investigation has led to some preliminary understanding \cite{PazSin}. One can
use this simpler theory of particle-field interaction to gain some insight
into the stochastic limit of quantum cosmology and quantum gravity. This was
one of the original motivations for this line of research undertaken
currently. An even more ambitious goal is to follow in detail the transition
from stochastic to quantum, including quantum particle aspects. We can use the
more \textquotedblleft tractable\textquotedblright\ problem here of
particle-field interactions to gain insight into the transition from
stochastic and quantum regimes of gravity in the so-called `bottom-up'
approach to quantum gravity \cite{GRhydro}. On this issue some interesting
results have been obtained \cite{CharisQSto,CRVphysica} based on quantum
Brownian motion (QBM) models. The significant nonlinearities of locally
interacting particles and fields in the present problem makes for a better
model to study this issue.

Here we take a modest step in this direction by applying our methods to a
uniformly accelerated charge in a scalar field. We choose a background field
$\varphi_{ext}$ such that $\ddot{z}^{\mu}\ddot{z}_{\mu}=-a^{2}=$
\textit{constant} is a solution to Eq. (\ref{Full semiclassical limit}) if we
ignore the radiation reaction terms. It can be immediately verified that
uniformly accelerating solutions satisfy $\left(  \dot{z}_{\mu}\ddot{z}%
^{2}+\dddot{z}_{\mu}\right)  |_{\ddot{z}^{2}=\text{constant}}=0$. Hence,
neglecting the suppressed higher derivative terms and quantum corrections, the
radiation reaction force identically vanishes, and the uniformly accelerating
trajectory is a solution to Eq. (\ref{Full semiclassical limit})
\emph{including} radiation reaction. (This results also holds if we use the
perturbative first derivative approximation to the semiclassical equations of motion.)

We choose a coordinate frame where the average trajectory is in the $\bar
{z}^{2}=\bar{z}^{3}=0$ plane, so that%
\begin{align}
\bar{z}^{0}\left(  \tau\right)   &  =a^{-1}\sinh\left(  a\tau\right)  ,\\
\bar{z}^{1}\left(  \tau\right)   &  =a^{-1}\cosh\left(  a\tau\right)  ,\\
\bar{z}^{2}\left(  \tau\right)   &  =\bar{z}^{3}\left(  \tau\right)  =0.
\end{align}
Defining right-moving coordinates $u\left(  \tau\right)  =\left(  \bar{z}%
^{1}-\bar{z}^{0}\right)  /2$ and left-moving coordinates $v\left(
\tau\right)  =\left(  \bar{z}^{1}+\bar{z}^{0}\right)  /2,$ the uniformly
accelerated trajectory is $u\left(  \tau\right)  =\left(  1/2a\right)
e^{-a\tau}$ and $v\left(  \tau\right)  =\left(  1/2a\right)  e^{+a\tau}.$ The
future particle horizon, shown in Fig. 1 (a), is given by $u=0.$%

\begin{figure}
[t]
\begin{center}
\includegraphics[
height=2.7674in,
width=5.7881in
]%
{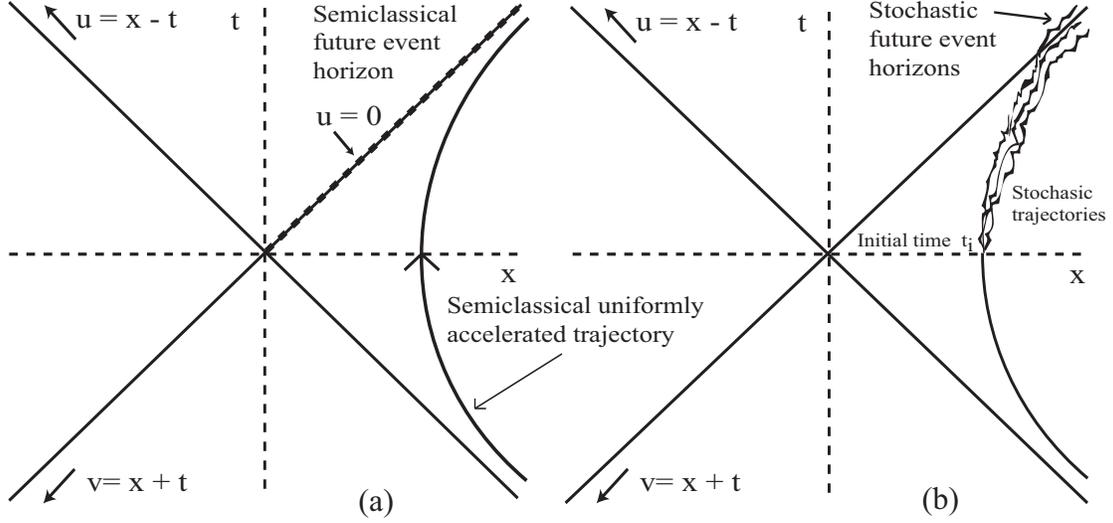}%
\caption{(a) Semiclassical trajectory of uniformly accelerated particle with
vanishing Abraham-Lorentz-Dirac (ALD) radiation reaction. Particle worldine in
far future ($\tau\rightarrow\infty$) gives future event horizon. (b)
Stochastic trajectories induced by quantum field fluctuations. Ensemble of
stochastic worldlines in far future give stochastic future even horizon. }%
\end{center}
\end{figure}

Let us now examine the linearized stochastic fluctuations of the particle,
induced by the stochastic force $\eta_{\mu}=e\vec{w}_{\mu}\tilde{\varphi
}\left(  \bar{z}\right)  ,$ around the semiclassical trajectory. The structure
of the field fluctuations are characterized by the anticorrelator (the
Hadamard function) along the accelerated trajectory:%
\begin{equation}
G_{H}\left(  \bar{z}\left(  \tau\right)  ,\bar{z}\left(  \tau^{\prime}\right)
\right)  =\left\langle \left\{  \hat{\varphi}\left(  \bar{z}\left(
\tau\right)  \right)  ,\hat{\varphi}\left(  \bar{z}\left(  \tau^{\prime
}\right)  \right)  \right\}  \right\rangle .
\end{equation}
For a massless scalar field in the vacuum state, evaluated at two different
times $\tau$ and $\tau^{\prime}$, the Hadamard function gives
\begin{align*}
G_{H}(\bar{z}\left(  \tau\right)  ,\bar{z}^{\prime}\left(  \tau\right)  )  &
=\text{Im}\left(  \frac{i}{4\pi^{2}}\int_{0}^{\infty}dk\ \frac{\sin
k\left\vert \bar{z}^{1}\left(  \tau\right)  -\bar{z}^{1}\left(  \tau^{\prime
}\right)  \right\vert }{\left\vert \bar{z}^{1}\left(  \tau\right)  -\bar
{z}^{1}\left(  \tau^{\prime}\right)  \right\vert }\exp\left(  k\bar{z}%
^{0}\left(  \tau\right)  -\bar{z}^{0}\left(  \tau^{\prime}\right)  \right)
\right) \\
&  =\text{Re}\left(  \int_{0}^{\infty}dk\frac{\sin k\left[  t+r\right]  -\sin
k\left[  t-r\right]  }{8\pi^{2}r\left(  \tau,\tau^{\prime}\right)  }\right)  ,
\end{align*}
where
\begin{equation}
r(\tau,\tau^{\prime})=\left\vert \bar{z}^{1}\left(  \tau\right)  -\bar{z}%
^{1}\left(  \tau^{\prime}\right)  \right\vert ,
\end{equation}
and
\begin{equation}
t(\tau,\tau^{\prime})=\left(  \bar{z}^{0}\left(  \tau\right)  -\bar{z}%
^{0}\left(  \tau^{\prime}\right)  \right)  .
\end{equation}
Substituting the uniformly accelerated solution, it becomes (re-inserting
factors of $c$)%
\begin{equation}
G^{H}\left(  \tau,\tau^{\prime}\right)  =\int_{0}^{\infty}\frac{dk}{k}g\left(
k,a\right)  \left\{  \coth\frac{c\pi k}{a}\cos k\left(  \tau-\tau^{\prime
}\right)  \right\}  , \label{Hadamard function uniform acceleration}%
\end{equation}
where the general form for $g\left(  k,a\right)  $ for a massive scalar field
in $D+1$ spacetime dimensions is given in \cite{Anglin93} as
\begin{equation}
g\left(  k,a\right)  =\frac{c^{2}k\sinh\left(  c\pi k/a\right)  }{2^{D-3}%
a\pi^{(D+3)/2}\Gamma\left(  \left(  D-1\right)  /2\right)  }\int_{0}^{\infty
}r^{D-2}K_{ick/a}\left(  c^{2}\sqrt{m^{2}+r^{2}}/a\right)  ^{2}dr.
\end{equation}
$K_{ik/a}$ is a modified Bessel function of imaginary order. The function
$g\left(  k,a\right)  $ is the effective spectral density of the scalar field
environment seen by the particle. In general it depends on the particle's
acceleration $a$ and mass $m.$ When $m=0,$ the integral can be evaluated in
closed form. For $D=3$ (the case of $3+1$ spacetime which is considered here),
the effective spectral density is independent of the particle state of
motion:
\begin{equation}
g\left(  k,a\right)  =\frac{k^{2}}{2c\pi^{2}}.
\end{equation}

The form of the Hadamard function in Eq.
(\ref{Hadamard function uniform acceleration}) is precisely the same as that
for a bath of harmonic oscillators with temperature $T=\hbar a/2\pi ck_{B},$
and spectral density given by $k^{2}/2c\pi^{2}$. Thus, the scalar field vacuum
along the accelerated particle trajectory looks like a thermal state, with
temperature proportional to the acceleration.

The correlator of the linearized stochastic force on the particle, given by
\begin{equation}
\left\langle \left\{  \eta^{\mu}\left(  \tau\right)  ,\eta^{\nu}\left(
\tau^{\prime}\right)  \right\}  \right\rangle \equiv e^{2}\vec{w}^{\mu}\left(
\bar{z}\left(  \tau\right)  \right)  \vec{w}^{\nu}\left(  \bar{z}\left(
\tau^{\prime}\right)  \right)  G_{H}^{\Omega}\left(  \bar{z}\left(
\tau\right)  ,\bar{z}\left(  \tau^{\prime}\right)  \right)  ,
\end{equation}
governs the \emph{response} of the particle to the quantum field fluctuations.
Note that $\vec{w}_{2}\left(  \bar{z}\left(  \tau\right)  \right)  =\vec
{w}_{3}\left(  \bar{z}\left(  \tau\right)  \right)  =0,$ and thus the induced
fluctuations act only in the $\left(  z^{0},z^{1}\right)  $ plane: $\eta
_{3}^{\mu}\left(  \tau\right)  =\eta_{4}^{\mu}\left(  \tau\right)  =0,$ in the
linearized response approximation. (This results does not carry over to the
nonlinear Langevin equation [Eq. (\ref{Nonlinear ALDL})] when higher order
effects are included). Because the stochastic force is given by $\eta_{\mu
}=e\vec{w}_{\mu}\tilde{\varphi}\left(  \bar{z}\right)  ,$ the response of the
particle depends on its state of motion through $\vec{w}_{\mu}\left(  \bar
{z}\right)  .$ While the underlying quantum field looks thermal along the
uniformly accelerated particle's average trajectory, the fluctuations that are
induced in the motion are complicated by non-isotropic and relativistic
effects. In fact these effects make this model more interesting. First, we
note that the Langevin equation describes the stochastic behavior of a
particle's worldline, depicting fluctuations in both the space \emph{and} time
coordinates. This makes this system a simple analog for quantum cosmology with
the stochasticity in time induced by quantum fluctuations.

Second, the particle motion is an interesting analog for stochastic black hole
horizon dynamics. If we view the particle trajectory in the far future as
effectively determining the horizon, the Langevin equations for $\tau
\rightarrow\infty$ then describes a stochastic distribution of horizon
positions [see Fig. 1(b)]. We note that most of the stochastic spread of the
particle trajectory (and hence the horizon) occurs for $\tau<c/a.$ When the
particle approaches the velocity of light, Lorentz time dilation suppresses
(i.e. slows) fluctuations from the perspective of external observers. The
analogy with black hole horizon formation suggests that most of the stochastic
spread in a black hole's horizon occurs in the early stages of its formation.
We are using stochastic relativistic particle dynamics as a guide to the
investigation of the horizon fluctuations in black hole dynamics.

\section{Discussions, Applications and Further Development}

The open system concept coupled with the coarse-grained effective action and
the influence functional techniques have been applied to quantum Brownian
motion \cite{QBM}, interacting quantum fields \cite{CGEA,Cooper94,Banff}, and
stochastic semiclassical gravity \cite{StoGraRev}. The new task we have
undertaken in recent years is the adoption of the worldline quantization
method to this existing framework for the treatment of relativistic charged
particles moving in a quantum field \cite{PRJDissertation,JH0,JH1}. In this
alternative formulation of QED the particle worldline structure is highly
efficient for describing the particle-like degrees of freedom (such as
spacetime position), while the field structure most easily describes processes
like radiation and pair creation. Here in the Discussions, we summarize the
special features and merits of this approach and then mention its applications
and extensions.

\subsection{Special Features of World-line Influence Functional Approach}

\subsubsection{Particle versus field formulation, fixed versus dynamic
background}

The contrasting paradigms of particles versus fields give very different
representations and descriptions for the same physical processes. The success
of quantum field theory for the past century has catapulted the field concept
to the forefront, with the particle interpreted as excitations of field
degrees of freedom. In recent years, the use of the quantum-mechanical path
integral in string theory has inspired a renewed effort towards
particle-centric quantum formalisms. The so-called worldline quantization
method--where the particle's spacetime coordinate $x^{\mu}\left(  \tau\right)
$ is quantized--is especially useful for calculating higher-loop processes in
non-dynamical classical background fields. Background field treatment is a
useful and well-defined method in conditions where the quantum fluctuations of
the fields are small. However, many problems involving the dynamical degrees
of freedom of the field require the inclusion of field dynamics beyond the
fixed-background approximation. A simple example is radiation reaction. More
examples are found in semiclassical gravity \cite{SCGravity}, where spacetime
playing the role of `particle' is treated classically while interacting with
the quantum fields .

The proper treatment of radiation-reaction requires, at the least, consistency
between the particle's (quantum) average trajectory and the mean-field
equations of motion. When one treats the particle as quantum mechanical, but
coupled only to the mean-field background (and not the quantum field
fluctuations) one is working within the semiclassical regime \cite{SCGravity}.
(When quantum fluctuations of the particle motion are sufficiently small, one
recovers the classical regime \cite{QBM}). In our program, we go beyond the
semiclassical approximation to include the influence of quantum field
fluctuations on the moving particle, thus reaching over to the stochastic
regime. To do this properly, we must begin with the full quantum theory for
both particle and field, since a particle can never fully decouple from the
field. Also, since most field degrees of freedom remain unobserved, we treat
the particle as a quantum open system with the field acting as an environment.
This is how nonequilibrium and stochastic mechanics ideas enter, and why the
self-consistent quantum particle evolution is aptly described by stochastic
quantum dynamics.

\subsubsection{Backreaction and Self-consistently Determined Trajectories}

One major improvement of our approach to the problem of moving charges in a
quantum field is the consideration of self-consistent backreaction of the
quantum field on the particle in the determination of its trajectory. We also
find that conceptual issues are easier to resolve if we deal with such
problems at four distinct levels: quantum, stochastic, semiclassical and
classical, as explained earlier. Confusion will arise when one mixes physical
processes of one level with the other without knowing their interconnections,
such as drawing the equivalence between radiation reaction with vacuum
fluctuations. Before summarizing our thoughts for processes under
nonequilibrium conditions, which cover most cases save a few special yet
important situations, such as uniform acceleration, let us remark that these
well-known cases are what we would call `test field' or prescribed
(trajectory) approximations. They are not self-consistently determined with
backreaction considerations. These cases are easier to study because they
possess some special symmetry, such as is present for the uniform acceleration
case (Rindler spacetime), inertial case (Minkowski), or the eternal black hole
spacetime. They are so legitimatized--meaning, physically relevant--only if
the backreaction of the field on the particle permits such solutions (which
may not be difficult to attain if the field strength is weak).
It is under these special conditions that a detector will detect thermality.
These very special conditions are the presuppositions of the Unruh and Hawking
effects. In more general situations the noise kernel is nonlocal, which means
that the noise in the detector is colored and temperature is no longer a
strictly viable concept.

\subsubsection{Quantum Origin of the ALD Equation}

Quantum field fluctuations, while responsible for the decoherence in the
quantum particle (system) leading to the emergent semiclassical and classical
behaviors, also impart an effective classical stochastic noise in the
equations of motion for the particle trajectory \cite{QBM}. These stochastic
differential equations feature colored noise that encodes the influence of
quantum fluctuations in the field . When there is sufficient decoherence,
these equations give excellent approximations to the particle motion. Even in
the case of weak decoherence they can provide a reasonable approximation to
the particle correlation functions. More accurate results can be obtained from
the n-particle-irreducible (nPI) closed-time-path effective action under a
large N approximation, which is a nice way to systemize the hierarchy of
correlation functions from the CGEA \cite{CH00}. This scheme has been applied
to stochastic gravity \cite{RVlargeN}.

\subsection{Applications and further developments}

Finally we want to connect our work with existing work in other areas, mention
applications of the present results to related problems, and show directions
where this new approach can stimulate further developments.

In terms of theory, the world-line quantization formalism has been applied to
QED, QCD and string theory \cite{WLQECD,string}. Our interest in this present
problem and its particular formulation stemmed partly from stochastic gravity
\cite{StoGraRev}, which is the stochastic generalization of semiclassical
gravity. The relation of semiclassical and stochastic regimes was made clear
in the investigation of quantum Brownian motion \cite{QBM}. More interesting
is the relation between the quantum and the stochastic regimes. Some work has
appeared addressing this issue \cite{CharisQSto,CRVphysica} but more insight
is needed. This is of particular interest in the `bottom-up' approach to
quantum gravity \cite{KinQG}: What features of quantum gravity can one detect
from the lower energy (mesoscopic) phenomena as manifested in stochastic
gravity? We don't have a theory of quantum gravity but we can learn from a
known and well-proven one such as QED. The investigations in this program
offers a concrete example of how a full quantum theory can yield stochastic
and semiclassical features. This could partially illuminate the common
pathways from stochastic semiclassical gravity to quantum gravity.

In terms of applications our results are useful for a range of problems, from
the consideration of the quantum aspects of beam physics (see, e.g.,
\cite{Chen}) to the consideration of possible schemes for the detection of
Unruh radiation, e.g., in linear accelerators or storage rings (note however
our critique for certain proposals of the former \cite{CapHR} and
misconceptions in the so-called `circular Unruh effect' \cite{CapHJ}). The
stochastic trajectory from the backreaction of quantum fields studied in the
last section is an interesting analog model to quantify black hole event
horizon fluctuations.

The new approach as illustrated in this report can also serve as a launching
pad for the exploration of nonequilibrium dynamics of charges undergoing
arbitrary (self-consistently determined, not prescribed) motions, strongly
correlated quantum systems, relativistic quantum kinetic and stochastic
theories. It is currently being generalized to curved spacetime (see also
\cite{BasZir02}) and applied to gravitational radiation reaction problems.
Recently a field-theoretical derivation of the MSTQW equation was given and a
new stochastic radiation reaction equation proposed \cite{GalHu,GalHuLin}.
This approach is also being applied to the study of quantum entanglement and
teleportation of relativistic detectors \cite{AlsMil} (see also \cite{RHA,RHK}%
) and environmental decoherence and entanglement \cite{YuEbe,SADH,ChouHu}
issues. \newline

\noindent\textbf{Acknowledgement} BLH wants to thank Professor Larry Horwitz
and other organizers of this conference for their hospitality. This research
is supported in part by NSF grant PHY03-00710

\end{document}